\documentclass[11pt,twoside]{article}
\usepackage{asp2010}

\resetcounters

\begin{document}

\title{Magnetic Fields on Cool Stars}
\author{Ansgar Reiners$^1$
\affil{$^1$Universit\"at G\"ottingen, Institut f\"ur Astrophysik, Friedrich-Hund-Platz 1, D-37077 G\"ottingen, Germany}
}

\begin{abstract}
  Magnetic fields are an important ingredient to cool star physics,
  and there is great interest in measuring fields and their geometry
  in order to understand stellar dynamos and their influence on star
  formation and stellar evolution. During the last few years, a large
  number of magnetic field measurements became available. Two main
  approaches are being followed to measure the Zeeman effect in cool
  stars; 1) the measurement of polarized light, for example to produce
  magnetic maps, and 2) the measurement of integrated Zeeman
  broadening to measure the average magnetic field strength on the
  stellar surface. This article briefly reviews the two methods and
  compares results between them that are now available for about a
  dozen M-type stars. It seems that we see a great variety of magnetic
  geometries and field strengths with typical average fields of a few
  kG in active M-type stars. The interpretation of geometries,
  however, has not yet led to a clear picture of magnetic dynamos and
  field configuration, and work is needed on more observational data
  but also on the fundamental understanding of our measurements.
\end{abstract}

\section{Introduction}

The Sun and cool stars are known to harbor magnetic fields leading to
all phenomena summarized under the term \emph{stellar activity}. It
may be debatable whether magnetic fields are actually the most
interesting aspect of cool star and solar physics
\citep[cp.][]{1985ARA&A..23..239M}, but it is certainly an exciting
field that brings together a large variety of physical mechanisms and
subtle analysis techniques. This makes it sometimes difficult to
interpret observational results and compare them to theoretical
expectations -- even if both are available, or even if one compares
observations achieved from different techniques.

A particularly interesting class of stars are cool stars of spectral
type M. Covering the mass spectrum between $\sim 0.6$ and
$0.1\,$M$_{\odot}$, M dwarfs are the most frequent type of
stars. Within this mass range, objects can have very different
physical properties. The very important transition from partly
convective (sun-like) to fully convective stars happens in the M dwarf
regime, probably around spectral type M3/M4. Furthermore, atmospheres
of M dwarfs can be very different and both molecules and dust gain
importance as the temperature drops toward late spectral types. 

In this article, I will concentrate on measurements of magnetic fields
on M dwarfs because most of the currently available measurements of
cool star magnetic fields are from M dwarfs, which is because sun-like
(field) stars tend to be less active (less rapidly rotating because of
shorter braking timescales) implying lower average magnetic fields
that are more difficult to detect.

\section{Technical aspects}

To measure the magnetic fields of stars, determination of spectral
line splitting due to the Zeeman effect is the most commonly used
method. In this article, I will not give a comprehensive overview of
the Zeeman technique but try to emphasize a few technical aspects that
are particularly important for the interpretation of the currently
available measurements. More comprehensive discussions and reviews on
magnetic fields and their measurement in cool stars are, e.g.,
\citet{1992A&ARv...4...35L, 1995ApJ...439..939V, 2005LNP...664..183M,
  2009ARA&A..47..333D, 2009AnRFM..41..317M}, or the conference reviews
by \citet{1996IAUS..176..237S, 2008ASPC..384..145J}.

\begin{figure}
\includegraphics[width=0.6\hsize]{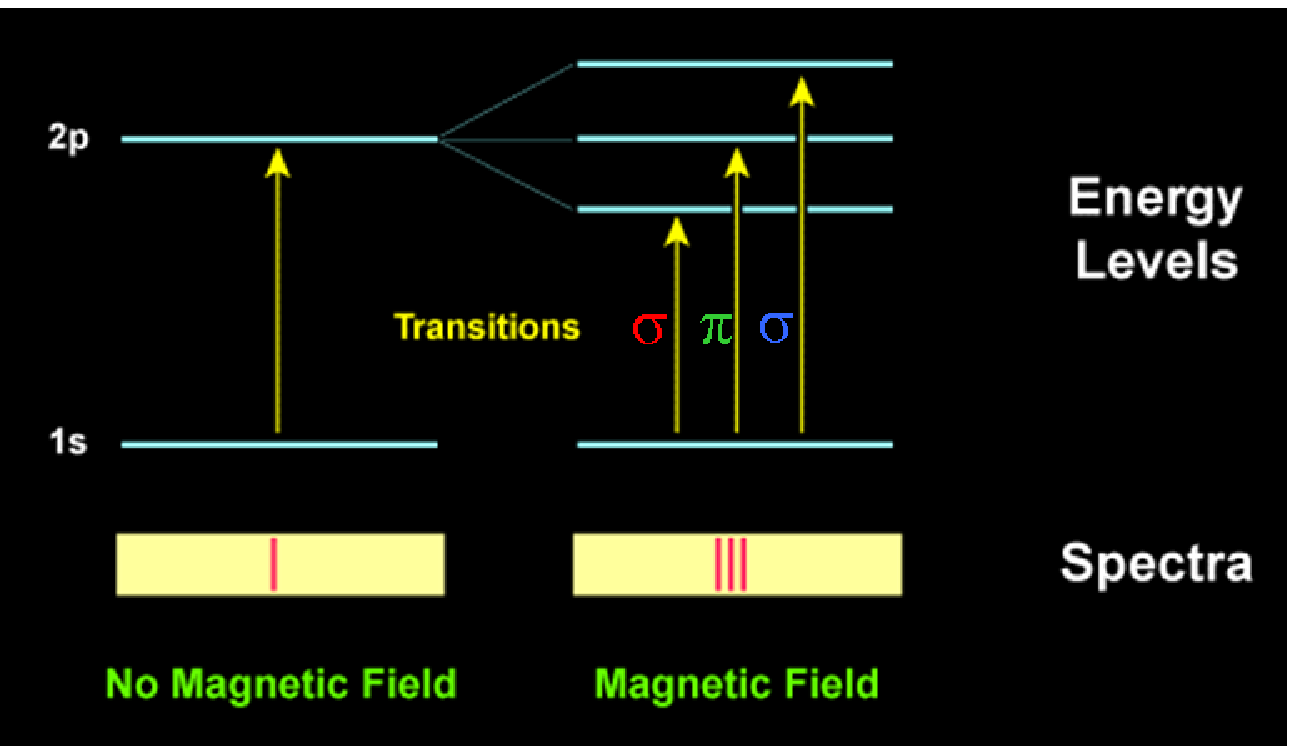}\includegraphics[width=0.4\hsize]{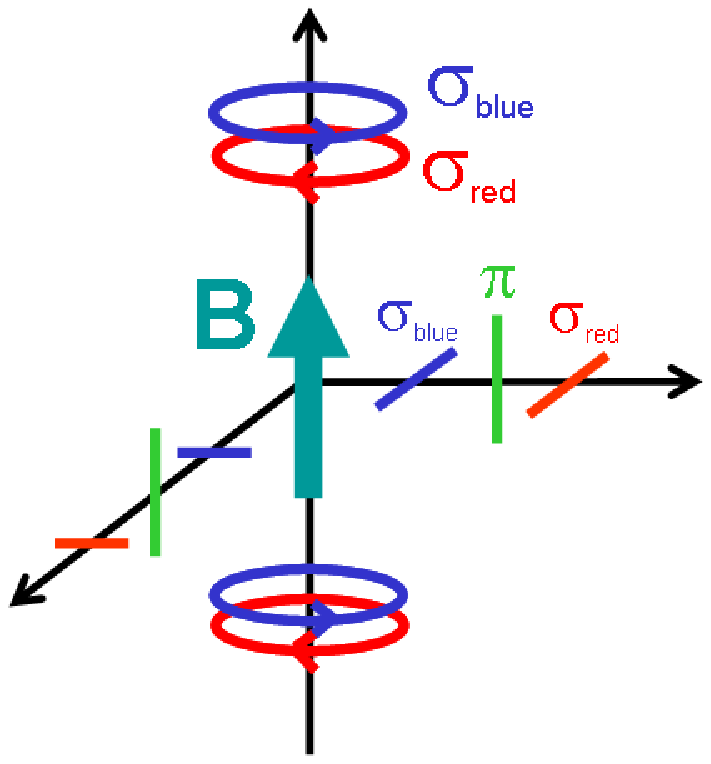}
\caption{\label{fig:scheme}Schematic view of Zeeman splitting. The
  upper level in the example (left panel) is split into three levels
  producing three spectral lines that are separated. The polarization
  of the $\pi$ and $\sigma$ components are shown in the right panel.}
\end{figure}

In short, dipole transitions obey the selection rule $\Delta M = -1,
0$ or $+1$. The three groups of lines according to different $\Delta
M$ are degenerate if no magnetic field is present, but separated in
the presence of a magnetic field. Transitions with $\Delta M = 0$ are
called $\pi$-components, $\Delta M = \pm1$ are called
$\sigma$-components. If a magnetic field is present, the energy
(wavelength) of the $\sigma$-components are shifted according to the
sensitivity of the transition (summarized in the \emph{Land\'e}-factor
$g$), the strength of the magnetic field $B$, and the wavelength of
the transition itself ($\lambda_0$). The average velocity displacement
of the spectral line components can be written as
\begin{equation}
  \label{Eq:Zeeman}
  \Delta v = 1.4 \, \lambda_0 g B,
\end{equation}
with $B$ in kG, $\lambda_0$ in $\mu$m, and $\Delta v$ in km\,s$^{-1}$.
Fig.\,\ref{fig:scheme} shows a simplified scheme of the splitting
(left panel) and of the polarization (right) of the $\pi$ and $\sigma$
components; the $\pi$ component is always linearly polarized while the
$\sigma$ components can be linearly or circularly polarized depening
on the viewing angle.

In order to measure a magnetic field, the measurement of different
polarization states can be of great advantage. This is immediately
clear from the right panel of Fig.\,\ref{fig:scheme} since the
different Zeeman components are polarized in a characteristic fashion.
A commonly used system are the Stokes components I, Q, U, and V, which
are defined in the following sense:
\begin{center}
\begin{tabular}{ccccc}
  I & = & $\updownarrow$ & $+$ & $\leftrightarrow$\\ 
  Q & = & $\updownarrow$ & $-$ & $\leftrightarrow$\\ 
  U & = & $\searrow \hspace{-3.8mm} \nwarrow$ & $-$ & $\swarrow \hspace{-3.8mm} \nearrow$\\ 
  V & = & $\circlearrowleft$ & $-$ & $\circlearrowright$\\ 
\end{tabular}
\end{center}
Stokes~I is just the integrated light (unpolarized light), Stokes~Q
and U measure the two directions of linear polarization, and Stokes~V
measures circular polarization. It has been shown that the magnetic
field distribution of a star can be reconstructed from observations of
all four Stokes parameters \citep{2002A&A...388..868K,
  2010arXiv1008.5115K}, and that using only a subset of Stokes vectors
leads to ambiguities that should be interpreted with caution.
Unfortunately, measurements of linear polarization are extremely
challenging in cool stars so that typically only Stokes~I and
(sometimes) Stokes~V are available.

\subsection{Field, flux, and filling factor}

The situation shown in Fig.\,\ref{fig:scheme} is simplified. In most
cases, line splitting is a bit more complicated, but the main
difficulty in measuring the splitting is the small value of $\Delta v$
compared to other broadening agents like intrinsic temperature and
pressure broadening, and rotational broadening. In a kG-magnetic
field, typical splitting at optical wavelengths is below
1\,km\,s$^{-1}$, which is well below intrinsic line-widths of several
km\,s$^{-1}$ and also below the spectral resolving power of typical
high-resolution spectrographs. Thus, individual components of a
spectral line can normally not be resolved even if the star only had
one well-defined magnetic field component. Real stars, however, can be
expected to harbor a magnetic field distribution that is much more
complex than this. Thus, even if spectral lines were intrinsically
very narrow and spectral resolving power infinitely high, we would
expect the Zeeman-broadened lines to look smeared out since in our
observations we integrate over all magnetic field components on the
entire visible hemisphere.

An important consequence of the fact that individual Zeeman-components
are usually not resolved is the degeneracy between magnetic field $B$
and filling factor $f$. A strong magnetic field covering a small
portion of the star looks similar to a weaker field covering a larger
portion of the star. An often used way around this ambiguity is to
specify the value $Bf$, i.e., the product of the magnetic field and
the filling factor (if more than one magnetic component is considered,
$Bf$ is the weighted sum over all components). Products of $B$ with
some power of $f$, for example $Bf^{0.8}$ are also considered because
they seem to be better defined by observations \citep[see, e.g.,][for
a deeper discussion]{1995ApJ...439..939V}. One important point to
observe is that $Bf$ is often called the ``flux'' -- because it is the
product of a magnetic field and an area -- but it has the unit of a
magnetic field. In fact, the term flux is very misleading since 1)
$Bf$ is identical to the average magnetic field on the stellar
surface, i.e., $Bf \equiv <$$B$$>$, and 2) the total magnetic flux of
two stars with the same values of $Bf$ can be extremely different
according to their radii because the actual flux is proportional to
the radius squared; $\mathcal{F} \propto Bfr^2$. As a consequence, the
value $Bf$ will be much lower in a young, contracting star compared to
an older (smaller) one if flux is conserved.

A related source of confusion is the difference between the signed
magnetic field (or flux), and the unsigned values or the square of the
fields (used to calculate magnetic energy). With Stokes~I, both
polarities produce the same signal and only the unsigned flux is
measured. This implies that Stokes~I carries only partial information
about field geometry, but it also means that Stokes~I always probes
the entire magnetic flux of the star. On the other hand, Stokes~V can
provide information on the sign of the magnetic fields, but this comes
with another serious caveat: Since we cannot resolve the stellar
surface, and Stokes~V measures the signed field, regions of opposite
magnetic fields on the stellar surface cancel out and can become
invisible to the Stokes~V signal.

\begin{figure}[!ht]
  \plotone{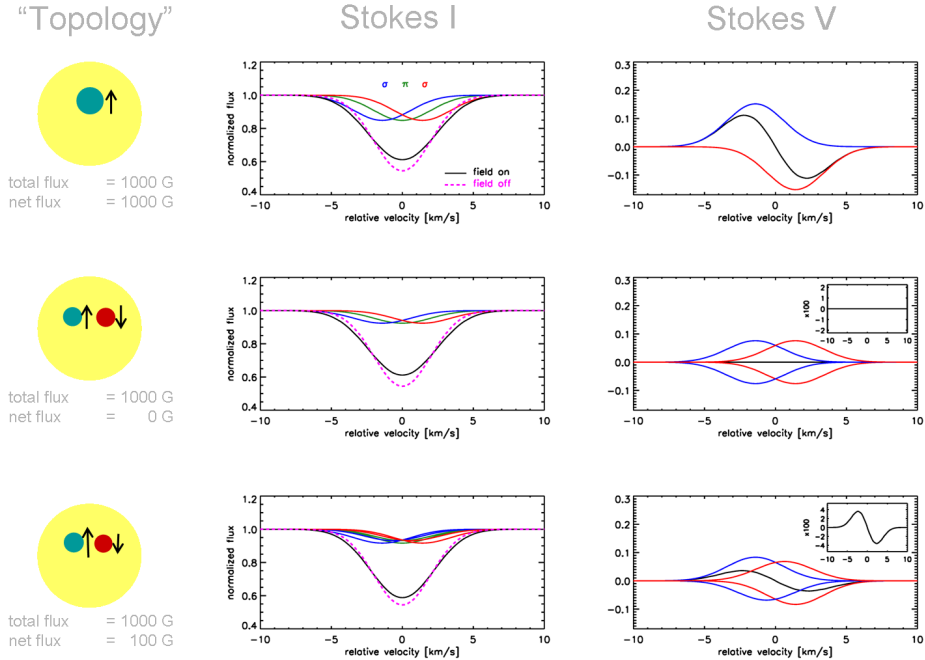}
  \caption{\label{fig:Stokes}Three examples of simplified field
    topologies and their signals in Stokes~I and V.}
\end{figure}

Examples of signatures of magnetic fields in Stokes~I and V are
sketched in Fig.\,\ref{fig:Stokes}. The left panel shows the
``topology'', which is actually not the topology of a stellar magnetic
field, but nothing else than two areas of radial magnetic fields put
on a flat surface (the spherical shape of a star has not been taken
into account in this example). In the top row, a simple magnetic field
region with only one direction is shown; signed and unsigned ``net''
flux both are 1~kG in this example. The line in Stokes~I is
effectively broadened, and Stokes~V shows a clear signal on the order
of 10\,\%. Note that the direction of the Stokes~V signal indicates
the direction of the magnetic field. In the second row, two magnetic
regions with only half the size as in the first example are
observed. Both regions have the same absolute field strength and area
but with opposite polarity. In this example, the Stokes~I signal is
identical to the first example, but the signal in Stokes~V entirely
vanishes because the net (signed) flux of this configuration is
exactly zero; any field strength in this cancelling configuration is
invisible to Stokes~V. The last row shows a case in which one of the
two areas is slightly larger than the other, the total flux is still
1000~G, but the net flux is only 100~G. The amplitude in Stokes~V is
ca. 4\,\%.

\subsection{Viewing angle}

Another thing that becomes immediately clear is that the geometric
interpretation of Zeeman splitting on an unresolved stellar disk can
be arbitrarily complicated, no matter if polarized or unpolarized
light is used. In addition to the ambiguity between magnetic field
strength and filling factor (which includes our ignorance about the
number and distribution of magnetic components), the signature of a
magnetic field region in stellar spectra depends on the angle between
the magnetic field lines and the line of sight
(Fig.\,\ref{fig:scheme}). In reality, again, a distribution of angles
will be present because field lines are probably bent on the stellar
surface, and because the stellar surface is spherical. As a result,
even geometrically relatively simple field distributions will lead to
complicated splitting patterns. If the star is rotating (as most stars
probably are) that pattern again depends a lot on the time a star is
observed -- which in turn can be utilized by observing the change of
observed spectra with rotation phase.

\section{Measurements of cool star magnetic fields}

Reliable detections of magnetic fields in cool stars of spectral type
F--K are relatively rare, see, e.g., \citet{1996IAUS..176..237S,
  2001ASPC..223..292S}. The challenges of detecting Zeeman splitting
in these stars using atomic lines at optical wavelengths have recently
been revisited by \citet{2010arXiv1008.2213A}. As realized earlier
\citep[e.g.,][]{1990ApJ...360..650B}, magnetic field signatures
can often be mimicked by the signatures of cool spots so that a
definite measurement of magnetic fields, and in particular their
distribution and geometry, is a delicate task.

M dwarfs, however, are a bit more cooperative. Although their spectra
exhibit a much denser forest of absorption lines rendering it
difficult to investigate isolated spectral lines, their magnetic
fields can be higher than fields in hotter stars at given rotational
velocity $v\,\sin{i}$. An important step in our understanding of M
dwarf magnetic fields was the investigation of seven M dwarfs by
\citet{1996ApJ...459L..95J}. Line shapes of an atomic Fe~I line at
846.7\,nm, hidden in a forest of TiO molecular lines, were analyzed
and average magnetic field strengths of several kG were reported in a
few stars that are probably fully convective (the results of their
analysis were updated by those reported in \citet{2000ASPC..198..371J}
using a multi-component approach). Complementary work looking for
magnetic field signatures in Stokes~V was done in T~Tauri stars very
successfully. Using polarized light, signatures of kG-strength
magnetic fields could be discovered
\citep[see][]{2007ApJ...664..975J}.

During the last years, two main routes have been followed searching
for direct detections of magnetic fields in M dwarfs. One method is to
employ near-infrared molecular absorption lines of FeH to search for
Zeeman broadening in Stokes~I, another is to extract the polarization
signal from several hundred lines at optical wavelengths in order to
search for field signatures in Stokes~V. Both methods proved to be
successful down to the latest M dwarfs, and it is possible to compare
the results of both methods in a small number of stars. In the
following, an overview on both methods is given and the results are
compared and interpreted.

\subsection{Results from Stokes~I}

The spectra of M dwarfs are dominated by the presence of molecular
absorption bands. At optical wavelengths, bands from TiO and VO
appear. Analysis of Zeeman broadening in these bands, however, is
difficult because the lines are not individually resolved. According
to Eq.\,\ref{Eq:Zeeman}, Zeeman broadening (in terms of velocity
shift) is proportional to wavelength, which implies that it is easier
to detect at longer wavelength, in particular because other Doppler
broadening agents like rotation do not depend on wavelength. At
near-infrared wavelengths, a molecular band of FeH can be found in the
spectra of M dwarfs, and \citet{1999asus.book.....W} showed that this
band is well suited for the measurement of magnetic fields. An
observational problem of FeH is that its most suitable band is located
at around $\lambda = 1\mu$m, which is too red for most CCDs and too
blue for most astronomically used infrared spectrographs. As a
consequence, only very few high-resolution spectrographs can provide
spectra of this wavelength, and the efficiencies are typically
ridiculously low. On the other hand, M dwarfs emit much of their flux
at near-infrared wavelengths so that in comparison to optical
wavelengths, the signal quality around 1$\mu$m is not much lower than
around 700\,nm if the spectra are obtained with an optical/near-IR
echelle spectrograph like HIRES (Keck observatory) or UVES (ESO
VLT). The 2004-upgrade of the HIRES spectrograph allowed a thorough
test of the FeH spectral range. We developed a method to
semi-empirically determine the magnetic fields of M dwarfs comparing
FeH spectra of our targets to spectra taken of two template stars; one
with no magnetic field and one with a known, strong magnetic field
\citep{2006ApJ...644..497R}. As reference, spectra of stars with
magnetic fields previously measured in a detailed analysis of atomic
lines are used \citep{2000ASPC..198..371J}. Thus, all magnetic field
measurements are relative to the reference star (Gl~873, $<B> =
3.9$\,kG), and magnetic fields higher than this value cannot be
quantified.

Obviously, systematic uncertainties of the measurements are quite
large, typically several hundred Gauss. Unfortunately, Zeeman
splitting of the FeH molecule is complicated and cannot entirely be
described so far \citep[see][]{2002A&A...385..701B}. Meanwhile,
progress has been made using an empirical approach to model Zeeman
splitting in FeH lines, and this approachs suggests that the fields
determined semi-empirically may be overestimated by some $\sim
20$\,\%\footnote{This could be due to an overestimate of the reference
  magnetic field measurement derived from the atomic line analysis.}
\citep{2010arXiv1008.2512S}.

\subsubsection{Average fields in M-type stars}

\begin{figure}[!ht]
  \plotfiddle{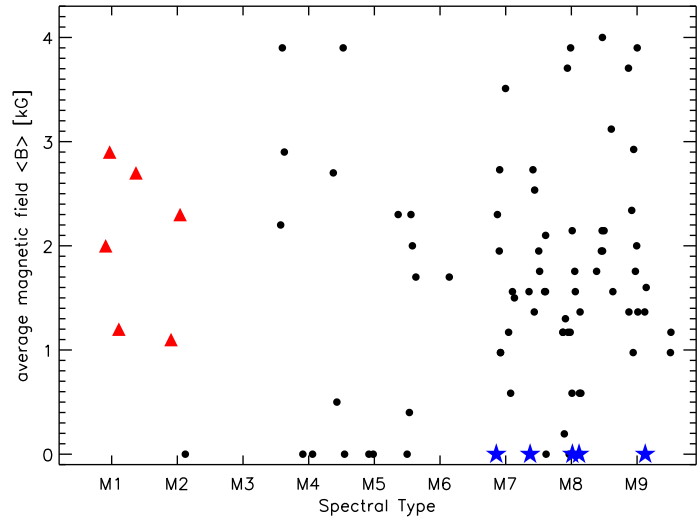}{.48\textwidth}{0}{45}{45}{-145}{0}
  \caption{\label{fig:BFields}Measurements of M dwarf magnetic fields
    from Stokes I. Red triangles: young, early M-stars; blue stars:
    young accreting brown dwarfs; black circles: field M dwarfs. See
    text for references. Spectral types are offset by a small amount
    to enhance visibility of different objects.}
\end{figure}

Magnetic field measurements in M dwarfs are reported in
\citet{1994IAUS..154..493S, 2007ApJ...664..975J, 2007ApJ...656.1121R,
  2009A&A...496..787R, 2010ApJ...710..924R, 2009ApJ...697..373R,
  2009ApJ...692..538R}. This is a non-exhaustive list of publications,
and a number of magnetic field measurements of stars re-analysed here
can be found in earlier literature (with more or less consistent
results); see \citet{1996IAUS..176..237S, 2001ASPC..223..292S}.
Measurements of M dwarf magnetic fields for spectral types M1--M9 are
shown in Fig.\,\ref{fig:BFields}.

It is well known that early-M dwarfs (M0--M3) in the field are much
less active and rotating slower than later, fully convective M stars
\citep[e.g.,][]{2008ApJ...684.1390R}. Early-M dwarfs appear to suffer
much more effective rotational braking so that their activity lifetime
is shorter than in later M dwarfs. Whether this is an effect of
different magnetic field topologies in partially and fully convective
stars is not clear -- this question is one of the basic motivations
for the comparison of field measurement results summarized here. It is
important to realize that at spectral type M3/M4, several parameters
of the stars change dramatically so that the reason for a change in
braking timescales may actually be much more fundamental than magnetic
field topology.

The field strengths of young, early-M and field mid- and late-M dwarfs
are on the order of a few kG. This is the main results from Zeeman
analysis and consistently found using different indicators (at least
in mid-M dwarfs). Compared to the Sun, the average magnetic field
hence is larger by two to three orders of magnitude, an observational
result that must have severe implications for our understanding of
low-mass stellar activity. It is not clear whether our picture of a
star with spots more or less distributed over the stellar surface is
actually valid in M dwarfs. If, for example, 50\,\% of the surface of
a star with a mean magnetic field of 4\,kG is covered with a ``quiet''
photosphere and low magnetic field, the other half of the star must
have a field strength as large as $\sim$8\,kG. The two components of
the stellar surface on such a star probably have very different
temperatures and properties, and the definition of effective
temperature must be considerably different from the temperature of the
``quiet'' photosphere.

In early-M dwarfs (M3 and earlier), magnetic fields were found in
young stars that are still rapidly rotating. Since old, early-M dwarfs
in the field are generally slowly rotating and inactive there has been
no search for magnetic fields in any large sample of them. Typical
field values can be expected to be on the level of a few hundred Gauss
and less, which is difficult to detect with Stokes~I Zeeman
measurements.

\subsubsection{Young stars and young brown dwarfs}

While young, early-M stars exhibit magnetic fields of several kG,
which is consistent with the magnetic fields of older M stars rotating
at comparable pace and young stars of earlier spectral type, it is
surprising that in young brown dwarfs of spectral types M7--M9 only
upper limits of a few hundred Gauss could be determined for their
magnetic fields \citep{2009ApJ...697..373R} (blue stars in
Figs.\,\ref{fig:BFields} and \ref{fig:Bvsini}). Interestingly, all
five young brown dwarfs seem to be accretors implying that they still
have a disk, and a magnetic field of a few hundred Gauss appears to be
enough for magnetospheric accretion in such a low-mass object. So far,
no direct magnetic field measurement could be performed in
non-accreting, old, field brown dwarfs, but it is expected that they
also harbor substantial magnetic fields \citep{2010A&A...522A..13R}.
Observations of radio-emission indicate that fields of a few kG
strength are in fact present on some L-type field brown dwarfs
\citep{2008ApJ...684..644H, 2009ApJ...695..310B}.

It is an open question whether the non-detection of magnetic fields in
brown dwarfs is due to the presence of accretion disks around the
objects observed so far. If this is the case, there ought to be some
mechanism for the disk to regulate the magnetic field of the central
object, which is not easily understood. Alternatively, large
difference in radius may be of importance in this context because the
surface area of young brown dwarfs is about an order of magnitude
larger than the surface of old brown dwarfs. If magnetic flux is
approximately conserved during its evolution, the average magnetic
field would be an order of magnitude lower in young, large brown
dwarfs than in old, small, field brown dwarfs.

\subsubsection{Rotation and magnetic fields}

\begin{figure}
  \plotfiddle{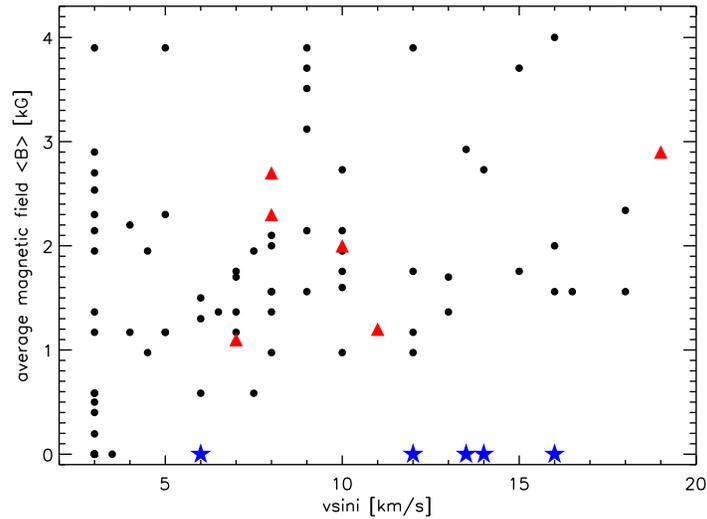}{.48\textwidth}{0}{45}{45}{-145}{0}
  \caption{\label{fig:Bvsini}Measurements of M dwarf magnetic fields
    from Stokes I as a function of projected rotation velocity,
    $v\,\sin{i}$. Symbols are the same as in Fig.\,\ref{fig:BFields}.}
\end{figure}

Stokes~I measurements of magnetic fields are shown as a function of
projected surface velocity $v\,\sin{i}$ in Fig.\,\ref{fig:Bvsini}. The
typical signature of a $\Gamma$-shaped rotation-activity relation is
visible, which means that the lower (left) end of the relation is not
resolved because the non-saturated part of the rotation-activity
relation falls below the detection limit in rotation velocities. On
the other hand, all rapidly rotating ($v\,\sin{i} \ga
3$\,km\,s$^{-1}$) field stars show detectable magnetic
fields. However, again, the young brown dwarfs violate this relation
since they are rapidly rotating but do not show detectable fields.

\begin{figure}
  \plotfiddle{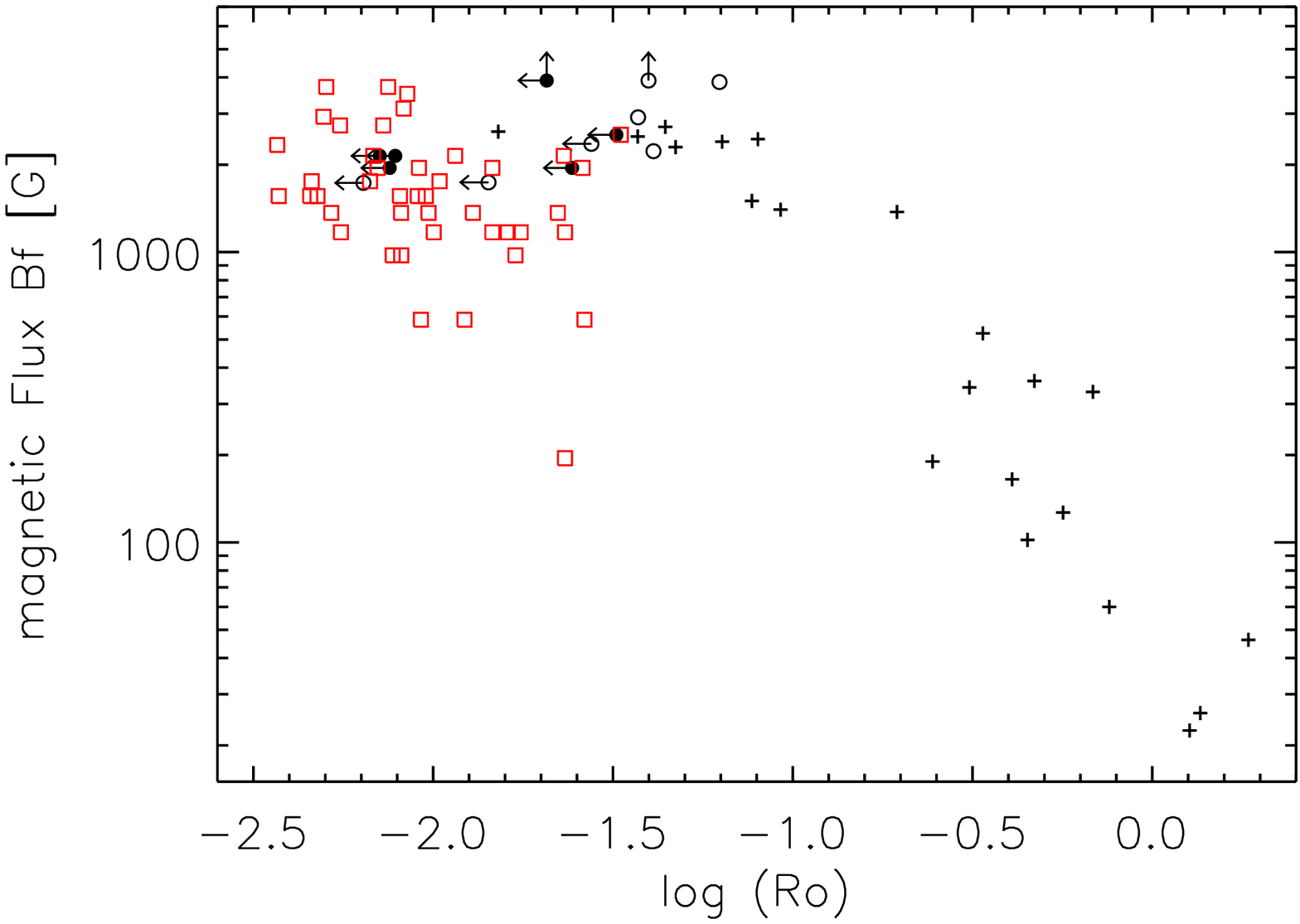}{.45\textwidth}{0}{40}{40}{-127}{0}
\caption{\label{fig:Bf_Ro}Magnetic fields as a function of Rossby
  number. Crosses and circles are stars M6 and earlier \citep[see][and
  references therein]{2009ApJ...692..538R}. Red squares are objects of
  spectral type M7--M9.}
\end{figure}

The rotation-activity relation describes the fact that slowly rotating
stars are less active than rapid rotators. In terms of Rossby number,
$Ro = P/\tau_{\rm{conv}}$, with $P$ the rotation period and
$\tau_{\rm{conv}}$ the convective overturn time, stars with $Ro \la
0.1$ are saturated in activity. Stars with larger values of $Ro$ show
activity proportional to $Ro$. We know from the Sun that activity is
caused by magnetic fields, and a rotation-magnetic field relation was
shown by \citet{1996IAUS..176..237S}. At that time, however, magnetic
fields could not be measured in stars with very low Rossby numbers
(saturated regime) because spectral line widths are too broad due to
the rotational broadening occuring at these velocities. M dwarfs, on
the other hand, have very small radii (and long overturn times) so
that for low Rossby numbers the corresponding surface velocities are
relatively low. This allows measuring magnetic fields of stars well
within the saturated regime. For M dwarfs of spectral type M6 and
earlier, it was found that average magnetic fields indeed saturate
\citep{2009ApJ...692..538R}. This implies that the saturation of the
rotation-activity relation is due to a saturation of the average
magnetic field and that $B$ itself is limited (in contrast to a limit
in the filling factor $f$ only).

Recently, we have measured magnetic fields in a sample of M7--M9 stars
\citep{2010ApJ...710..924R}. These measurements are shown as red
squares in Fig.\,\ref{fig:Bf_Ro}. Stars as cool as M7 seem to deviate
from the rotation-activity relation; they still show higher magnetic
fields at lower $Ro$, but saturation seems to occur at much lower
values of $Ro$.

\subsection{Results from Stokes~V}

Maps of magnetic fields in M stars derived from time-series of
Stokes~V measurements are presented in \citet{2008MNRAS.390..545D,
  2008MNRAS.390..567M, 2010MNRAS.407.2269M}. The Stokes~V profiles are
derived simultaneously from several hundred lines through
least-squares-deconvolution (LSD). LSD makes use of the weak-field
approximation so that very strong magnetic fields (several kG) may not
be detectable with this method. As mentioned above, the use of only
Stokes~V means that the magnetic geometry will not be fully
characterized. In particular, magnetic regions of opposite polarity
can cancel each other so that they remain undetected. A way to
overcome this issue is to take observations at different times so that
magnetic regions may remain visible when they appear at the limb of
the star (Zeeman Doppler Imaging, ZDI). The resolution of such a
Doppler image critically depends on the rotation velocity of the star
and the exposure time required to obtain the necessary data quality.

A tremendous amount of work was put into the analysis of magnetic
geometries in stars through ZDI, and the possibility of reconstructing
magnetic fields on stellar surfaces is truly amazing. However, the
interpretation of the field maps is very difficult, and conclusions
have to be drawn with great care.

\subsubsection{Magnetic field strengths}

\begin{figure}
\plotone{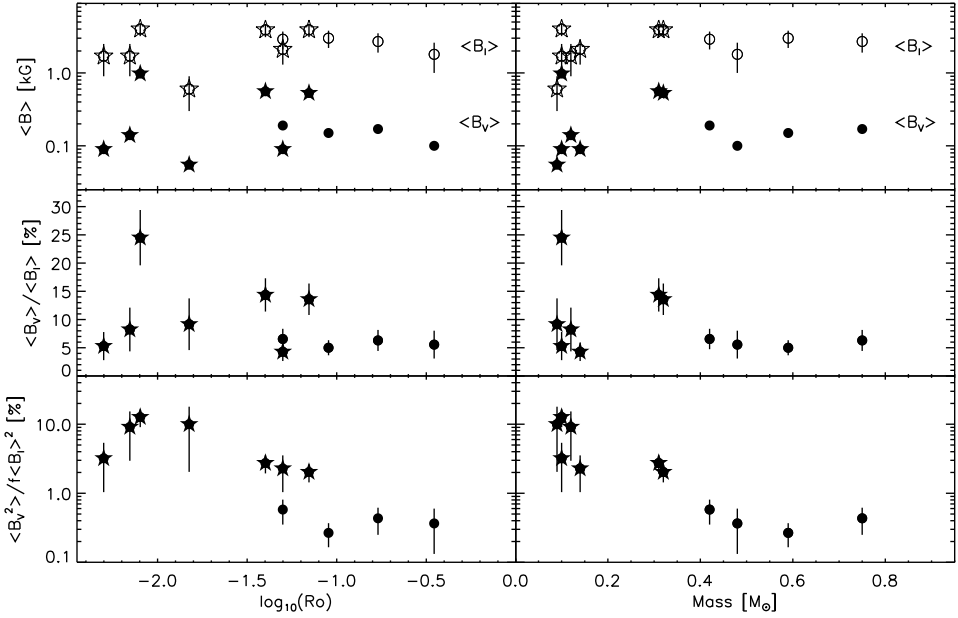}
\caption{\label{fig:BigPlot}Measurements of M dwarf magnetic fields
  from Stokes~I and Stokes~V. \emph{Top panel:} Average magnetic field
  -- Open symbols: measurements from Stokes~I; Filled symbols:
  measurements from Stokes~V. \emph{Center panel:} Ratio between
  Stokes~V and Stokes~I measurements. \emph{Bottom panel:} Ratio
  between magnetic energies detected in Stokes~V and Stokes~I. Circles
  show objects more massive than 0.4\,M$_{\odot}$, stars show objects
  less massive than that. }
\end{figure}

Typical average magnetic field strengths found in Stokes~V
measurements of M dwarfs are about a few hundred Gauss. Note that this
is the average value for the detected unsigned magnetic field, $|B|$,
the same as in Stokes~I measurements; the average value of the signed
magnetic field is zero by construction. An average field strength of a
few hundred Gauss is much lower than average field strengths of
magnetically active stars observed in Stokes~I that are a typically
few kG. The literature today contains eleven M dwarfs for which
magnetic field measurements were carried out independently both in
Stokes~V and Stokes~I. I compare the results from the mentioned papers
in Fig.~\,\ref{fig:BigPlot}, which is an update of Fig.\,2 in
\citet{2009A&A...496..787R}.

Fig.\,\ref{fig:BigPlot} shows the average magnetic fields from
Stokes~I and V, their ratios, and the ratios of magnetic energies as a
function of Rossby number and stellar mass. In the top panel, the
measurements are shown directly, the center panel shows the ratio
between the average magnetic fields $<$$B_V$$>$$/$$<$$B_I$$>$. For the
majority of stars, the ratio is on the order of ten percent or less,
which means that $<$10\,\% of the full magnetic field is detected in
the Stokes~V map. In other words, more than 90\,\% of the field is
invisible to this method. As discussed above, this is probably a
consequence of cancellation between field components of different
polarity. One notable exeption is the M6 star WX\,Uma, which has an
average field of approximately 1\,kG in Stokes~V (Gl~51 shows an even
higher field but has not yet been investigated with the Stokes~I
method).

A second observable that comes out of the Stokes~V maps is the average
squared magnetic field, $<$$B^2$$>$, which is proportional to the
magnetic energy of the star. Under some basic assumptions, this value
can be approximated from the Stokes~I measurement, too
\citep[see][]{2009A&A...496..787R}. The ratio between approximate
magnetic energies detected in Stokes V and I is shown in the bottom
panel of Fig.\,\ref{fig:BigPlot}, it is between 0.3 and 10\,\% for the
stars considered.

\subsubsection{Topologies of the detected fields}

In contrast to the conclusions suggested in
\citet{2008MNRAS.390..545D} and \citet{2009A&A...496..787R}, evidence
for a change is magnetic topologies at the boundary between partial
and complete convection is not very obvious when the new results are
included. Four of the late-M dwarfs have ratios
$<$$B_V$$>$/$<$$B_I$$>$ below 10\,\% while earlier results suggested
that more flux is detectably in Stokes~V in fully convective stars. On
the other hand, the ratio of detectable magnetic energies stays rather
high in this regime ($\ga 2\%$), which may reflect an influence of the
convective nature of the star. The main problem, however, seems to be
why some stars have very different ratios
$<$$B^2$$>$/$<$$B$$>^2$. This may well be an effect of different
magnetic topologies but large differences occur even within the group
of fully convective stars \citep[see][]{2010MNRAS.407.2269M}.

\section{Summary}

Our knowledge on magnetic fields in cool stars, particularly in M
stars across the full convection boundary, has seen enormous progress
during the last few years. Intensive observations of many M dwarfs led
to the construction of Stokes~V Doppler maps, and the exploitation of
the FeH molecular spectra allow a determination of the entire field
from Stokes~I. We can now start to compare results from independent
methods and search for the influence of stellar parameters including
convective nature. The interpretation of results from different
methods opens a parameter space that certainly contains deep
information about the fields and their topology, but it is not yet
clear what our measurements are actually telling us. Field strengths
and topologies have ramifications to a broad range of astrophysics,
and at spectral type late-M, we are approaching the brown dwarf
regime. More field measurements, determination of molecular constants,
and fundamental investigation of detectabilities are required to push
the field forward, and to understand the many facets of magnetic
fields in cool stars, brown dwarfs, star formation, and their links to
exoplanets.

\acknowledgements 

It is a pleasure to thank my main collaborators in the work on
Stokes~I magnetic fields, Gibor Basri, Denis Shulyak, and Andreas
Seifahrt, and I thank Julien Morin for insightful discussions. I want
to thank the organizers of CS16 for a very fruitful and extremely
well-organized meeting. My work is funded through a DFG Emmy-Noether
fellowship under RE~1664/4-1.

\bibliography{reiners_a}

\end{document}